# Review of Robust Video Watermarking Algorithms

Mrs Neeta Deshpande

Research Scholar
STRM University
Nanded India
e-mail: Deshpande_neeta@yahoo.com

Dr. Archana rajurkar

Professor and Head
MGM College of Engineering
Nanded India
e-mail: Archana_rajurkar@yahoo.com

Dr. R. manthalkar

Professor and Head
SGGS Institute of Engineering and Technology
Nanded India
e-mail: rmanthalkar@yahoo.com

*Abstract*—**There has been a remarkable increase in the data exchange over web and the widespread use of digital media. As a result, multimedia data transfers also had a boost up. The mounting interest with reference to digital watermarking throughout the last decade is certainly due to the increase in the need of copyright protection of digital content. This is also enhanced due to commercial prospective. Applications of video watermarking in copy control, broadcast monitoring, finger printing, video authentication, copyright protection etc is immensely rising.**

**The main aspects of information hiding are capacity, security and robustness. Capacity deals with the amount of information that can be hidden. The skill of anyone detecting the information is security and robustness refers to the resistance to modification of the cover content before concealed information is destroyed. Video watermarking algorithms normally prefers robustness. In a robust algorithm it is not possible to eliminate the watermark without rigorous degradation of the cover content. In this paper, we introduce the notion of Video Watermarking and the features required to design a robust watermarked video for a valuable application. We review several algorithms, and introduce frequently used key techniques. The aim of this paper is to focus on the various domains of video watermarking techniques. The majority of the reviewed methods based on video watermarking emphasize on the notion of robustness of the algorithm**

**Keywords-Video Watermarking, Authentication, Robust, Techniques, DCT, DWT.**

## I. INTRODUCTION

Video Watermarking is a young and rapidly evolving field in the area of multimedia. Following factors have contributed towards the triggering of interest in this field. a) The society is contaminated by the tremendous piracy of digital data, as copying of digital media has become comparatively easy.
b) This is an era where need has arise for fight against "Intellectual property rights infringements".
c) Copyright protection must not be eroded due to malicious attacks
d) Tampering of the digital data needs to be concealed at some point.

The requirement of secure communication and digital data transfer has potentially increased with the development of multimedia systems. Data integrity is not secure in image transfers [16]. The main technique used for protection of Intellectual Property rights and copyright protection is digital watermarking [1, 2, and 3]. The copyright data may be in the form of text [4] [5], image [6]-[9], audio [10], video [11-15]. Watermarking may be visible or invisible. Invisible watermarking implies that the presence of the watermark is barely discernible when the watermarked signal is displayed. Watermark embedding may bring in diminutive distortion into the audible or visible components of the watermarked signal. If the watermark cannot be easily removed from the watermarked signal even after applying common watermarking attacks then it is referred as robust embedding.

The basic components involved in robust watermarking are *watermark embedding*, *attack*, and *watermark detection*. In watermark embedding, a watermark signal (Text, image or audio etc) is constructed and then embedded into an original signal (Video in context with this paper) to produce the watermarked signal. Once embedding is done, the watermarked video can be subjected to various attacks. During watermark detection, the watermark detector is given a test signal that may be watermarked, attacked or not. The watermark detector reports whether the watermark is present or not on examining the signal at its input.

According to the grapevine, any image watermarking technique can be extended to watermark videos, but reality speaks that video watermarking techniques need to meet other challenges than that in image watermarking schemes. Basically, the watermarking technologies are classified in various domains like the spatial domain or temporal domain [17] [18] and the frequency domain [19-22].

This paper is organized six sections. The subsequent section explains the important aspects of video watermarking. Section III focuses the widespread applications of video watermarking .Section IV considers the robustness aspect by elaborating on the common attacks in video watermarking. The various domains of video watermarking are explored and a robust algorithm in each domain is considered in Section V.

This is a vast section as has six main subsections dealing with spread spectrum domain, Spatial, DCT, DWT, Feature Domain and SVD domain of video watermarking. And a conclusion is presented in the last section.

## II.    IMPORTANT ASPECTS OF VIDEO WATERMARKING

Video watermarking embeds data in the video for the purpose of identification, annotation and copyright. A number of video watermarking techniques have been proposed. These techniques exploit different ways in order to embed a robust watermark and to maintain original video fidelity. Conventional encryption algorithms permit only authorized users to access encrypted digital data. Once such data are decrypted, however, there is no way in prohibiting its illegal copying and distribution.

Many algorithms for developing watermarks on images are extended for videos. But some points need to be considered during the extensions.

  a)  Between the frames there exists a huge amount of intrinsically redundant data.
  b)  There must be a strong balance between the motion and the motionless regions
  c)  Strong concern must be put forth on real time and streaming video applications

The following aspects are important for the design of video watermarking systems.

  a)  Imperceptibility: The watermark embedding should cause as little degradation to the host video as possible.
  b)  Robustness: The watermark must be robust to common signal processing manipulations and attempts to remove or impair the watermark.
  c)  Security: The embedded information must be secure against tampering.
  d)  Capacity: The amount of embedded information must be large enough to uniquely identify the owner of the video

## III.    APPLICATIONS OF VIDEO WATERMARKING

The present section is completely devoted to the presentation of various applications in which digital watermarking can convey a precious hold up in the perspective of video. Digital video watermarking is used in a variety of applications. Various applications are elucidated in detail in [23]. Some major ones are

  a)  Fingerprinting: In this technique the video is uniquely identified by its resultant fingerprint by software that recognizes extracts and then compresses distinguishing components of a video.  Some of the features that are involved in video fingerprinting analysis are key frame analysis, color changes, motion changes etc. of a video sequence. In this technique watermarks are embedded as fingerprints on the video. Several fingerprinting methods extract the fingerprints on the video. The evaluation and identification of the video content is then performed by comparing extracted fingerprints. The white paper [24] provides a high level overview of digital watermarking and fingerprinting and examines how these two technologies can be integrated into workflows for automatically tracking, protecting and monetizing content.

  b)  Copy control: Copy protection is a widely exercised application in video watermarking. In this a watermark is used to indicate whether a video content is copyrighted. This watermark can only be removed with a severe degradation of the video sequence. The work on copy protection issues in DVD is being carried out by The Copy Protection Technical Working Group (CPTWG). However a system, which could become the future specification for DVD copy protection, has been defined [25]. various issues that play a role in designing a copy-protection system for digital versatile disk (DVD) video as perceived by Millennium, one of the two contenders in the DVD-video copy protection standardization activity are being addressed in [26].

  c)  Broadcast Monitoring: In broadcast monitoring the content owner embeds the watermark prior to transmission. The watermark is extracted by the monitoring site that is set up within the transmission area. The central database server receives this information that can be used for additional supplementary applications like transmission verification, IPR surveillance etc. A video watermarking technology for broadcast monitoring. Is focused in [27].

  d)  Video Authentication: In applications involving instance videos captured by surveillance cameras, checking the integrity of the images and the video is a major issue. Fragile, semi fragile and robust watermarking are the commonly used policies. A slight modification on the cover video destroys fragile watermarks. Semi fragile watermarking can resist content conserving operations and be sensitive to content varying transforms. In [28] a semi fragile watermarking scheme is proposed that can be applied on images and video.  This scheme uses entropy of the probability distribution of gray level values in block groups to generate a binary feature mask, which is embedded robustly into an adjacent I-frame. The scheme is robust to content-preserving manipulations and sensitive to content-changing manipulations.

  e)  Copyright protection: copyright protection of video data is an important issue in digital video delivery networks. There are many techniques of video watermarking for copyright protection. In one of the techniques a watermark is added to the video signal that carries information about sender and receiver of the delivered video. In [29] a novel self-adaptation  differential energy watermarking based on the Watson  visual model is proposed that inserts robust watermark into video streaming according to the differential energy theory.  A comprehensive approach for protecting and managing video copyrights in the Internet with watermarking techniques is presented in [30].

## IV. COMMON ATTACKS IN VIDEO WATERMARKING

As with any encryption/ secret encoding scheme, the algorithms used are subject to attacks. Watermarking is no exception and many techniques have been derived to effectively attack watermarked systems. The common attacks of video watermarking are frame dropping, frame averaging, statistical analysis, lossy compression, cropping and various signal processing and geometrical attacks.

In [31] the author has classified the attacks on video watermarking basically in two categories
Intentional attacks: The intentional watermark attack include Single frame attacks like filtering attacks, contrast and color enhancement and noise adding attack. Or statistical attacks like averaging attack and collision attack.

Unintentional attacks: The unintentional attacks may be due to Degradations that can occur during lossy copying, or due to Compression of the video during re encoding or because of Change of frame rate and Change of resolution

## V TECHNIQUES IN VIDEO WATERMARKING

Many algorithms have been proposed in the scientific literature for robust watermark embedding in video. We explore some most commonly used domains used for video watermarking and explore one algorithm in each domain.

a) Spread Spectrum Domain video watermarking Technique: Scheme 1
One of the pioneer works in video watermarking considers the video signal as a one dimensional signal [32]. The algorithm for watermarking in the uncompressed domain mentioned in this paper for embedding watermark can be summarized as follows
Watermark embedding algorithm

1. A one dimensional signal is acquired by a simple line scanning. Let the sequence $x_j \in \{-1, 1\}$ represents the

Watermark bits to be embedded. This sequence is spread by a chip-rate $cr$ according to the following equation:

$$y_i = x_j \quad j.c_r \leq i < (j+1).c_r \quad i \in N$$

2. Embedding one bit of information into $cr$ samples of the video signal by spreading operation adds redundancy.

3. The obtained sequence $y_i$ is then amplified locally by an amplitude factor $\beta(i) \geq 0$ and modulated by a pseudo-random binary sequence $p_i \in \{-1,1\}$ $i \in N$

4. The modulated spread spectrum watermark is obtained by

$$w(i) = \beta(i) * y_i * p_i \quad i \in N$$

5. The spread spectrum watermark $w(i)$ is added to the line-scanned video signal $v(i)$, which gives the watermarked video signal $p_i$

$$W_{vi}' = v_i + \beta(i) * y_i * p_i \quad i \in N$$

The adjustable factor $\beta(i)$ may be tuned according to local properties of the video signal, e.g. spatial and temporal masking of he HVS, or kept constant depending on the targeted application.
Watermark Extraction: The watermark recovery is done by correlation.

1. The watermarked video sequence $Wvi'$ is high-pass filtered, resulting in a filtered watermarked video signal $v''$ so that major components of the video signal itself are isolated and removed.

2. Demodulation is carried out. The filtered watermarked video signal is multiplied by the pseudo-random noise $p(i)$ used for embedding and summed over the window for each embedded bit. The correlation sum $s(j)$ for the $j$th bit is given by the following equation:

$$s_j = \sum_{i=j.cr}^{(j+1).cr-1} p_i.v'' = \sum_{i=j.cr}^{(j+1).cr-1} p_i.Wv_i' + \sum_{i=j.cr}^{(j+1).cr-1} p_i.p_i.\beta_i.y_i$$

3. The correlation consists of two terms $\Sigma_1$ and $\Sigma_2$. The main purpose of filtering was to leave $\Sigma_2$ untouched while reducing $\Sigma_1$ down to 0. As a result, the correlation sum becomes:

$$s_j = \sum 1 + \sum 2 \approx \sum_{i=j.cr}^{(j+1).cr-1} p_i^2.\beta_i.y_i. = x_{j.\sigma_p^2}.cr.mean(\beta_i)$$

4. The hidden bit is then directly given by the sign of $s(j)$
This pioneer method proposes a very flexible framework, which can be used as a basic root of a more complicated video watermarking scheme.

b) Spatial Domain video watermarking Technique: Scheme2
We begin our exploration of video watermarking techniques in the spatial domain, also referred to as the pixel or coordinate domain. Algorithms in this class generally share the following characteristics:
1. The watermark design and the watermark insertion procedures do not involve any transforms.
2. Simple techniques like addition or replacement are used for the combination of watermark with the host signal and embedding takes place directly in the pixel domain. The main strengths of pixel domain methods are that they are conceptually simple and have very low computational complexities. As a result they have proven to be most attractive for video watermarking applications where real-time performance is a primary concern. Macq and Quisquater [33] focused on the issue of watermarking digital images. In the paper, the authors illustrated a technique to insert a watermark into the LSB of pixels that are located in the vicinity of image contours. As the LSB technique was implied, modifications of

LSB's destroyed the watermark However, the LSB techniques also exhibit some major limitations

- Since absolute spatial synchronization is required, susceptibility to de-synchronization attacks is increased
- Multiple frame collusions may occur due to lack of consideration of the temporal axis.
- Watermark optimization is difficult using only spatial analysis techniques.

A video watermarking with robustness against rotation, scaling and translation (RST) is proposed in [34]. The watermark information is embedded into pixels along the temporal axis within a Watermark Minimum Segment (WMS). Since the RST operations for every frame along the time axis in video sequence are the same at a very short interval, the watermark information can be detected from watermarked frames in each WMS subjected to RST. The algorithm presented in this paper has a basic assumption that the geometrical transformation for every frame along the time axis in a video sequence is the same at a very short interval. So the position of each pixel Watermarking Minimum Segment (WMS) is changed in same way along the time axis.

The algorithm in the paper can be summarized as

Before watermarking, a time-axis template, which carries the information of at least three original pixel positions, is embedded into the video sequence along the time-axis within each WMS. In the algorithm a random sequence $v_i(k)$ is defined as the time axis template having the following property

$$< v_i, v_j > = \delta_{ij} \quad v_i \quad ^2$$

Embedding Algorithm

1. An original video sequence is first segmented into its WMS
2. Template pixel positions are defined pseudo randomly by a secret key in each WMS.
3. The time-axis template is embedded in the spatial-domain of each frame as described in 1. in the following way

$$S'_k(x, y) = S_k(x, y) + \beta(\sigma_{x,y}(k)+1)vi(k) \quad ----1$$

Where $S_k(x, y)$ is the original pixel value of the frame, $S'_k(x,y)$ is the modified pixel value of the frame, $(x, y)$ is the pixel position of the time-axis template embedded, $\beta$ is a global scaling factor, $\sigma_{x,y}$ is the standard deviation of the pixel values along the time axis in each WMS with length of $N$, and $k=0,1,2,…N-1$.

Recovery Algorithm:

1. The number of templates needed for recovery depends on the quality of the recovery frame for geometrical distortions. Generally 3 templates are enough for watermark recovery.
2. The time-axis template can easily be detected through

$$< S'_{k-l}(x', y') * h.v_i(k) > = d \quad ------2$$

Where $h$ is a prediction filter,* is convolution operator, $l$=0, 1, 2…N-1, and $N$ is the length of time-axis template along the time axis. The variable $l$ is used to find the start point of the time-axis template embedded in the video frames. If $d$ >threshold the time-axis template is detected, and the new pixel positions $(x', y')$ can be attained.

The paper further proposes to embed a watermark in the same way as the template was embedded. For the generation of the watermark, the author has taken 30 PN sequences that are representative of 90 message bits of W1 in WMS time of about 1 second and 9 PN sequences for W2 in WMS time of 5 sec.

So for embedding equation 1 is modified to

$$S'_k(x, y) = S_k(x, y) + \beta(\sigma_{x,y}(k)+1)W_i(k) \quad ------3$$

Here $W_i(k)$ is the watermark that is coded by the technique of direct sequence code division multiple access. (DS_CDMA) [35][36].

The recovery of the watermark is given by

$$d_{xy}(i) = 1/M \sum_{k=i.M}^{i+1-M} (S'_{k-1}(x,y) * h).v_{(i)}(k) \quad -------4$$

And

$$d_{xy} = \sum_{i=0}^{[N/M]} \left| d_{xy}(i) \right| \quad ----------------5$$

Where $h$ is a prediction filter, * is convolution operator, [·] is integer operator, $l$=0, 1, 2…$N$-1, $N$ is the length of WMS or the length of embedded PN sequence, and $M$ (=Tb/Tc) is a bandwidth expansion factor. The variable $l$ is used to find the start point of embedded sequence in the video frames. If $dxy$ >threshold, the watermark is detected, and

$$b'_i = \begin{cases} +1 & if \ d_{xy}(i) > 0 \\ -1 & if \ d_{xy}(i) < 0 \end{cases}$$

The author further solves the problem of synchronization along the time axis. This is done by embedding the special reference orthogonal sequences $ri(k)$ with the same length of WMS multiple times at different points along the time axis before watermarking.

So the embedding algorithm becomes

$$S'_k(x+q._{x,y}+q._y) = S_k(x+q._{x,y}+q._y) + \beta(\sigma_{x+q._{x,y}+q._y}(k)+1)W_i(k)$$

where $k=0,1,2,…N-1$, $N$ is the length of WMS, $q$ is representative of embedding times ($q$=0,1 in the presented algorithm), and $i$=1,2.

The detection is based on autocorrelation given by

$$R_{x,y}(u,v) = < S'_k(x,y) * h.F'_k(x+u, y+v) * h >$$

Where $h$ is a prediction filter, * is convolution operator.

Experimental results show that the technique is robust against the attacks of RST, bending and shearing of frames, MPEG-2 lossy compression, colour-space conversion, and frame dropping attacks.

c)  DCT domain Video watermarking technique: Scheme 3

The watermark signal is not only designed in the spatial domain, but sometimes also in a transform domain like the full-image discrete cosine transform (DCT) domain or block-wise DCT domain. Features of DCT

a)  The Characteristics of DCT coefficients must utilize few coefficients for providing excellent signal approximations.

b)  Since the frequency components are ordered in a sequential order, starting with low frequency, mid frequency and high frequency components, a proper selection of the components can be prepared.

c)  A smooth block is represented, if most of the high frequency coefficients are zero.

d)  An edge block is represented, if the low frequency coefficients have large absolute values.

DCT is faster and can be implemented in O (n log n) operations. The DCT allows an image to be broken up into different frequency bands, making it much easier to embed watermarking information into the middle frequency bands of an image. The middle frequency bands are chosen such that they avoid the most visual important parts of the image (low frequencies) without over-exposing themselves to removal through compression and noise attacks (high Frequency). The DCT transforms a signal or image from the spatial domain to the frequency domain.

In [37] the author proposed a framework based on the Hartung technique which depended on spread spectrum communication in discrete cosine transform (DCT). The basic algorithm used in this paper is designed by Hartung , Girod[38]. The basic principal is borrowed from spread spectrum. The algorithm mentioned in the paper can be summarized as

Embedding algorithm

1. For every frame of the video sequence find its DCT coefficients

A] A one dimensional signal is acquired by a simple line scanning. Let the sequence $x_j \in \{-1, 1\}$ represents the

Watermark bits to be embedded. This sequence is spread by a chip-rate $cr$ according to the following equation:

$$y_i = x_j \quad j.c_r \le i < (j+1).c_r \quad i \in N$$

B] Embedding one bit of information into $cr$ samples of the video signal by spreading operation adds redundancy.

C] The obtained sequence $y_i$ is then amplified locally by an amplitude factor $\beta(i) \ge 0$ and modulated by a pseudo-random binary sequence $p_i \in \{-1,1\}$ $i \in N$

D] The modulated spread spectrum watermark is obtained by

$$w(i) = \beta(i) * y_i * p_i \quad i \in N$$

E] The watermark bits are transformed by using the DCT.

Wi=DCT (Wi)

F] The watermark wi is directly added to the video signal vi

$$vi' = vi + wi \quad vi' = vi + wi$$

The watermark recovery is done by correlation.

1.  The watermarked video sequence $Wvi'$ is high-pass filtered, resulting in a filtered watermarked video signal $v''$ so that major components of the video signal itself are isolated and removed.

2.  Demodulation is carried out. The filtered watermarked video signal is multiplied by the pseudo-random noise $p(i)$ used for embedding and summed over the window for each embedded bit. The correlation sum $s(j)$ for the $j$th bit is given by the following equation:

$$s_j = \sum_{i=j.cr}^{(j+1).cr-1} p_i.v'' = \sum_{i=j.cr}^{(j+1).cr-1} p_i.Wv_i' + \sum_{i=j.cr}^{(j+1).cr-1} p_i.p_i.\beta_i.y_i$$

3.  The correlation consists of two terms $\Sigma_1$ and $\Sigma_2$. The main purpose of filtering was to leave $\Sigma_2$ untouched while reducing $\Sigma_1$ down to 0. As a result, the correlation sum becomes:

$$s_j = \sum 1 + \sum 2 \approx \sum_{i=j.cr}^{(j+1).cr-1} p_i^2.\beta_i.y_i = x_{j.\sigma_p^2}.cr.mean(\beta_i)$$

4.  The hidden bit is then directly given by the sign of $s(j)$

As per the results presented in the paper the scheme was robust to geometrical attacks like cropping, scaling and rotation in the DCT domain

d)  DWT domain Video watermarking techniques: Scheme 4

Another possible domain for watermark embedding is that of the wavelet domain. The DWT (Discrete Wavelet Transform) separates an image into a lower resolution approximation image (LL) as well as horizontal (HL), vertical (LH) and diagonal (HH) detail components. The process can then be repeated to compute multiple "scale" wavelet decomposition. One of the many advantages over the wavelet transform is that that it is believed to more accurately model aspects of the HVS as compared to the FFT or DCT. This allows us to use higher energy watermarks in regions that the HVS is known to be less sensitive to, such as the high resolution 32 detail bands LH, HL, HH). Embedding watermarks in these regions allow us to increase the robustness of our watermark, at little to no additional impact on image quality.

In [39] the author proposes a novel hybrid digital video watermarking scheme based on the scene change analysis and error correction code. The basic technique used for embedding the watermark is the Discrete Wavelet Transform..

The basic embedding algorithm in the paper can be summarized as

1. Watermark preprocess: - A watermark is scrambled into small parts as a part of preprocess. The watermark is first scaled to a particular size as

$$2 \wedge n <= m; n > 0 \quad \text{----------- (1)}$$

$$p + q = n; p, q > 0 \quad \text{----------- (2)}$$

Where m – No of scene changes and p,q,n – No of positive integer.
Size of watermark is determined by,

$$64.2 \wedge p * 64.2 \wedge q \quad \text{----------- (3)}$$

Then the watermark is divided into $2 \wedge n$ small images with size 64.

In the next step, each small image is decomposed into 8 bit-planes, and a large image can be obtained by placing the bit-planes side by side only consisting of 0s and 1s. These processed images are used as watermarks.

2. Video preprocess: -

All frames of the video are decomposed in 4-level sub band frames by separable two-dimensional (2-D) wavelet transform.

Scene changes are detected from the video by applying the histogram difference method on the video stream.

Independent watermarks are embedded in frames of different scenes.

3. Watermark embedding: The watermark is then embedded to the video frames by changing position of some DWT coefficients with the following condition:

*if Wj= 1 then*
*Exchange(Ci,Ci+1,Ci+2,Ci+3,Ci+4);*
*else*
*Exchange(Ci,Ci+1,Ci+2,Ci+3,Ci+4);*
*end if*

Where Ci is the ith DWT coefficient of a video frame, and Wj is the jth pixel of a corresponding watermark image [40].

4. watermark detection

The video is processed to detect the video watermark.

The detection is done by the following logic

*If (WC(i) > median (WCi,WCi+!,WCi+2,WCi+3,WCi+4);*
*Then EWj*
*Else EWj = 0*
*end if*

In the paper the author has also embedded an audio watermark in the audio sequence. The mentioned algorithm was experimented for attacks like Frame dropping, Frame collision, Cropping, Rescaling, Noise, Mpeg-attack, Lossy compression, Median filter, Row –column removal, Rotation and Affine

5. Feature Domain PCA based video watermarking technique : Scheme 5

The mathematical procedure of transforming a number of possibly correlated variables into a smaller number of uncorrelated variables is called Principal component analysis (PCA). The smaller numbers of uncorrelated variables are called *principal components*. Given a data set, the principal component analysis reduces the dimensionality of the data set. In [41] the author proposes a new digital video watermarking scheme based on Principal Component Analysis. The video shots are detected based on informational content, and color similarities. The key frames of each shot are extracted and each key frame is composed of three color channels. Embedding of the watermark is done in the three color channels RGB of an input video file.

The following PCA procedure adapted from [42] is explored in the watermarking process.

PCA process:

a) a) Assuming the data set as a vector

$$z = (z1, z2...zn)^{T} \quad (1)$$

b) Find the mean of the population

$$\mu_{z} = E\{z\} \quad (2)$$

c) Calculate the covariance matrix of the data set

$$c_{z} = E\{(z - \mu_{z})(z - \mu_{z})^{T}\} \quad (3)$$

The components of $c_{z}$, denoted by $c_{ij}$, represent the co variances between the random variable components $z_{i}$ and $z_{j}$.

d) Suppose $Z = (Z1, Z2...Zn)$ where (Z1,Z2…Zn) are sample vectors, calculate sample mean and the sample covariance matrix .

e) Calculate the eigen values $e_{i}$ and the corresponding eigen vectors $\lambda_{i}$ from the covariance matrix.

$$c_{z}e_{i} = \lambda_{i}e_{i} \quad (4)$$

f) A data set that has the most significant values can be found by creating an ordered orthogonal basis by ordering the eigen vectors in the descending sequence of the eigen values. This matrix $[e_{i}]$ is also called PCA basis function created by the eigen vectors $(e_{1}, e_{2}, e_{3}...e_{n})$.

g) The original vector Z can be deco related by the basis matrix $[e_{i}]$. By transforming data vector z

$$y = e_{i}(z - \mu_{z}) \quad (5)$$

The components of y are the coordinates in the orthogonal vector

h) The original data vector z can be recovered from $y$ using

$$z = e_{i}^{T} y + \mu_{z} \quad (6)$$

By selecting the eigen vectors having largest eigen values a very little information is lost in terms of mean square error thus preserving a varying amount of energy in the original data.

Embedding process

1. Decompose the video stream to sequences, then to scenes then to shots and then extract each frame in each shot, using key frame extraction technique in [43] based on spatio-temporal features of the shots.

2. Watermark is embedded in the key frames.

3. For each key frame, Separate the frame in 3 Sub frames by separating the R,G,B channels.

4. Apply PCA procedure explained above for for each sub frame .The principal components of each of the frames *FR*, *FG* and *FB* are computed then have the three PCA coefficients: *YR*, *YG*, and *YB*.

5. Watermark is a random signal that consists of a pseudo-random sequence of length M. *W = w1, w2...wM*. The watermark was embedded into the predefined components of each PCA sub-block uncorrelated coefficients.

$$(y_i)w = y_i + \alpha |y_i| w_i$$

Where $\alpha$ is a strength parameter

6. The three RGB watermarked color channels are separately recovered by the inverse PCA process and a watermarked frame is achieved

7. The video is reconstructed by retrieving first the video shots by the method in [44]

Extraction Procedure: The watermark is recovered by applying the correlation formula for each frame separately.

$$(CV) = WY * /M = 1/M \sum_{i=1}^{M} w_i y_i *$$

The preliminary results in the paper showed a high robustness against most common video attacks, especially frame cropping, cropping and recalling for a good perceptual quality.

6. SVD Domain Video Watermarking technique: scheme 6

Singular Value Decomposition (SVD) is a numerical technique for diagonal zing matrices in which the transformed domain consists of basis states that is optimal in some sense. The SVD of an N x N matrix A is defined by the operation:

$$A = U S V^T$$

Where *U* and *V* ∈ *R N x N* are unitary and *S* ∈ *R N x N* is a diagonal matrix. The diagonal entries of *S* are called the singular values of A and are assumed to be arranged in decreasing order $\sigma i = \sigma i + 1$. The columns of the *U* matrix are called the left singular vectors while the columns of the V matrix are called the right singular vectors of *A*. Each singular value $\sigma i$ specifies the luminance of an image layer while the corresponding pair of singular vectors specifies the geometry of the image layer [45, 46, and 47].

In [48] the author proposes two effective, robust and imperceptible video watermarking algorithms based on the algebraic transform of Singular Value Decomposition (SVD). In this paper we explore only the first one. In the mentioned algorithm, watermark bit information is embedded in the SVD-transformed video in a diagonal-wise fashion.

The algorithm mentioned in the paper can be summarized as Embedding Algorithm

1: The video clip is divided into video scenes $v_{si}$

2: the frames in each video scene are processed by SVD

3: Every video frame *F is converted to* YCBCR color matrix format from RGB.

4: The SVD for the Y matrix in each frame *F* is computed. This process generates 3 Matrices (U, S, and V) such as:

$$Y = U_Y S_Y V_Y$$

5: The size, of the watermark must match the size of the matrix which will be used for embedding U, V or S. So a rescaling is performed on the watermark image

6: Embedding can be done in one of the three SVD matrices: U, V, or S, as

A] Embedding in Matrix U Diagonal-wise

1. Each diagonal value $(u_{i,i})$ in The U matrix is inversed, such as $x = 1/(u_{i,i})$.

2. The binary bits of the watermark $W_{vsi}$ are embedded into the integer part of x by substituting the watermark bit $W_i$ with the 7th bit of x.

3. $u_{i,i'} = 1/x'$ is obtained by applying inverse to each x.

4. The modified coefficient matrix $U'$ *is* applied with inverse SVD such as:

$$y' = U_{y'} S_y V y^T$$

B] Embedding in Matrix V Diagonal-wise

1. Each diagonal value $(v_{i,i})$ in The V matrix is inversed, such as $x = 1/(v_{i,i})$

2. The binary bits of the watermark $W_{vsi}$ are embedded into the integer part of x by substituting the watermark bit $W_i$ with the 7th bit of *x*.

3. The modified values of V matrix, are obtained by applying inverse such that $v_{i,i'} = 1/x'$

4. Inverse SVD is applied on the modified coefficient matrix V '. Such as:

$$y' = U_y S_y V y^{'T}$$

C] Embedding in Matrix S Diagonal-wise

1. The binary bits of the watermark $W_{vsi}$ are embedded into the integer part of each diagonal value of the S matrix, $(s_{i,i})$ by substituting the watermark bit $W_i$ with the 7th bit of $(s_{i,i})$

2. Inverse SVD is applied on the modified coefficient matrix S '. Such as:

$$y' = U_y S_y 'V y^T$$

$y'$ is the updated luminance in the YCBCR color scheme.

And thus a final watermarked Video frame $F$ is obtained

7: The video frames are converted to RGB color matrix from YCBCR

8: The frames are reconstructed to form watermarked Video scene $v_{si'}$ .

9: The watermarked scenes are reconstructed to get the final watermarked Video clip.

Extraction Algorithm

1: The watermarked Video clip $V'$ is converted into watermarked scenes $v_{si'}$ . $Wvsi$

2: In each scene, the watermarked frames are processed using SVD

3: The video frame $F'$ are converted from YCBCR to RGB color matrix.

4: SVD for the Y matrix in frame $F'$ is calculated and 3 matrices are generated (U, S, V).

5: Extraction is done in one of the three SVD matrices: $U$, $V$, or $S$, in the following way

   A] Extraction from Matrix U

      1. Each diagonal value $(u_{i,i})$ in the U matrix is inversed such as $x = 1 / (u_{i,i})$

      2. The embedded watermark is extracted from the integer part of x

$$Wvsi(i) = 7thLSB(fix(X))$$

   B] Extraction from Matrix V

      1. Each diagonal value $(v_{i,i})$ in the Vmatrix is inversed, such as $x = 1 / (v_{i,i})$

      2. The embedded watermark is extracted from the integer part of x:

$$Wvsi(i) = 7thLSB(fix(X))$$

   C] Extraction from Matrix S

$$Wvsi(i) = 7thLSB(fix(Si,i))$$

6: The image watermark $Wvsi$ is constructed by cascading all watermark bits extracted from all frames.

7: The same procedure is followed for all video scenes.

    The results obtained in the paper produced high robustness against attacks such as frame compression, rotation, and frame dropping, frame swapping and frame averaging attacks.

## VI CONCLUSION

In the paper we revised various video watermarking algorithms proposed in the literature in various domains. A comparative study focusing on the robustness factor of each algorithm is mentioned in the Table 1.

New approaches are expected to come out and may merge existing approaches. For example, a watermark can be separated into two parts: one for copyright protection and the other for customer fingerprinting. One watermark may be embedded in DCT domain while other in DWT. various scene change algorithms may be suggested. Further texture features could be extracted. However many challenges have to be taken up. Robustness is a parameter that has to be well thought of. Some aggressive video processing's may modify the watermark signal. Certain factored checks have to be described for an intended application. Some major challenges are the collusion attack and real time watermarking as proposed in the literature. The performance of many image watermarking algorithms is being improved by the perceptual measures. It is challenging to exploit the perceptual features of the video in real time.

## AUTHORS PROFILE

Authors

Mrs Neeta Deshpande has secured her BE and ME in computer science and Engineering. Currently she is a research scholar in SRTM university Nanded India. She is working towards her PhD in the area of video watermarking. She has a teaching experience of twelve years in the field of computer engineering. She has published several papers in national and international conferences. Her research interests include Watermarking and Artificial intelligence.

Dr Archana rajurakar: Archana M Rajurkar received degree of BE ( Computer engineering ) and ME (Instrumentation) from Marathwada university Aurangabad and Ph.D (Computer science and Engineering) from IIT Roorkee, India. She joined as a faculty member in M.G.M.'s College of Engineering, Nanded in 1991. Currently She is working as Professor and Head in the Department of Computer Science and Engineering, M.G.M.' College of Engineering, Nanded, India. Her research interests include Content-Based Image and Video Retrieval, Multimedia and Image Databases, Computer Vision and Pattern Recognition

Dr R. Manthalkar: Dr manthalkar has completed his BE in 88 and ME in 94 from SGGS Nanded. He persued his PHD from IIT Kharagpur India in 2003. Currently he is working as an professor and head of the electronics department in SGGS institute of engineering and technology Nanded India.. His research interests include texture analysis, VLSI design and biomedical signal processing.

| Scheme No. | Domain | Category of watermark Visible or invisible | Cover | Preprocessing Of watermark | Frames selection | Robust to attacks | Recovery |
|---|---|---|---|---|---|---|---|
| 1 | Spread spectrum [32] | invisible | Bit sequence | Yes---spread sequence watermark | No | Block-wise Compression, Addition of Gaussian and impulse noise, Low pass filtering | Yes |
| 2 | Spatial Domain [ 34] | invisible | PN sequence | Yes | No | Rotating with cropping & scaling, symmetric and asymmetric shearing, Mpeg-2 attacks, random frame dropping | yes |
| 3 | Frequency Domain DCT Based [37] | invisible | Bit Sequence | Spread spectrum using chip_rate | No | Compression, Rotation, Scaling, Gaussian noise attack and salt and pepper attack | Yes |
| 4 | Frequency Domain DWT Based [39] | Invisible | Black and white image | Yes | Yes Scene based | Frame dropping, Frame collision, Cropping, Rescaling, Noise, Mpeg-attack, Lossy compression, Median filter, Row –column removal, Rotation , Affine | Yes |
| 5 | Feature Domain PCA based [41] | invisible | pseudo-random sequence | N0 | No | Cropping, Rescaling, Frame dropping, Rotation, Median Filter, | Yes |
| 6 | SVD Domain [48] | invisible | Image | Yes—Rescaling | Yes Scene based | JPEG compression, video frame rotation, noise attacks (Gaussian, and salt and pepper noise), frame Dropping and frame swapping & averaging. | Yes |

TABLE 1 SUMMARY OF THE REVIEWED VIDEO WATERMARKING ALGORITHMS